\documentclass[twocolumn,nofootinbib,amsmath,amssymb,aps,prd,balancelastpage]{revtex4-1}

\usepackage{graphicx}
\usepackage[caption=false]{subfig}

\usepackage{amsmath,amssymb}
\usepackage{amsfonts,amssymb,mathrsfs}

\usepackage[bookmarks=false,pdfstartview=FitH]{hyperref}
\usepackage[all]{hypcap}

\def\be{\begin{equation}}
\def\ee{\end{equation}}
\def\nn{\nonumber}
\def\f{\frac}

\def\pl{{\rm Pl}}
\def\lp{\ell_\pl}
\def\b{\bar}

\def\h{\hat}

\def\wh{\widehat}

\def\bra{\langle}
\def\ket{\rangle}
\def\dd{{\rm d}}

\def\de{\delta}

\def\vp{\varphi}

\usepackage{color}

\begin{document}

\pagestyle{plain}

\title{Bouncing cosmologies from quantum gravity condensates}

\author{Daniele Oriti} % \email{daniele.oriti@aei.mpg.de}
\affiliation{Max Planck Institute for Gravitational Physics (Albert Einstein Institute),\\
Am M\"uhlenberg 1, 14476 Golm, Germany, EU}

\author{Lorenzo Sindoni} % \email{lorenzo.sindoni@aei.mpg.de}
\affiliation{Max Planck Institute for Gravitational Physics (Albert Einstein Institute),\\
Am M\"uhlenberg 1, 14476 Golm, Germany, EU}

\author{Edward Wilson-Ewing} % \email{wilson-ewing@aei.mpg.de}
\affiliation{Max Planck Institute for Gravitational Physics (Albert Einstein Institute),\\
Am M\"uhlenberg 1, 14476 Golm, Germany, EU}

\begin{abstract}

We show how the large-scale cosmological dynamics can be obtained from the hydrodynamics of isotropic group field theory condensate states in the Gross-Pitaevskii approximation.  The correct Friedmann equations are recovered in the classical limit for some choices of the parameters in the action for the group field theory, and quantum gravity corrections arise in the high-curvature regime causing a bounce which generically resolves the big-bang and big-crunch singularities.

\end{abstract}

\maketitle

\section{Introduction}
\label{s.intro}

A major challenge for any theory of quantum gravity is to extract its physics at cosmological scales and make predictions about the early universe.

We will consider this problem within the scope of loop quantum gravity (LQG) \cite{Thiemann:2007zz} and group field theory (GFT) \cite{GFT}, in which the fundamental building blocks of space-time are quanta of geometry, elementary excitations above a fully degenerate `no-space' vacuum. The recovery of semi-classical gravity then requires the manipulation of states involving a large number of these elementary quanta, whose collective behaviour will determine the effective theory via coarse-graining of the microscopic dynamics.  For the case of the simplest example, the spatially flat, homogeneous and isotropic Friedmann--Lema\^itre--Robertson--Walker (FLRW) space-time, the calculation of the effective dynamics as determined by the particular quantum gravity model will require: the identification of a family of cosmological states, controlling the relevant global degrees of freedom (e.g., Hubble rate, matter energy density), and then a coarse-graining of the dynamics to extract the large-scale effective Friedmann equations.

Interestingly, even though loop quantum cosmology (LQC)---where symmetry-reduced space-times are quantized mimicking the procedures of LQG \cite{Ashtekar:2011ni}---starts from a different perspective, it gives further insight into the type of states that correspond to the cosmological sector of LQG.  While the exact relation between LQG and LQC remains  an open question, a heuristic relation between the two theories has been proposed \cite{Ashtekar:2006wn, Bojowald:2007ra, Ashtekar:2009vc, Pawlowski:2014nfa}, suggesting that the LQG states corresponding to the cosmological sector are spin network states with a large number of nodes, all labeled by the same quantum numbers, and where all links are coloured by the same $SU(2)$ representation label (typically taken to be $j = 1/2$), and the connectivity information of the nodes is disregarded.  Due to homogeneity, it is usually further assumed that all of the nodes in the spin network have the same number of links, e.g., 4.  In short, LQC suggests to consider states in which all the (many) quanta of geometry are in the same state, i.e., a condensate of quantum geometry.

All these features are naturally encoded in the family of GFT condensate states that have been recently used to extract cosmology from the full GFT formalism \cite{Gielen:2013kla, Gielen:2013naa, CondensatesReview}. GFTs are a field theory reformulation of LQG and spin foam models, and hence provide powerful tools that are well-suited to study condensate states of quanta of geometry.  We refer to \cite{Oriti:2016qtz} and references therein for a complete presentation of the conceptual and technical aspects related to GFT condensates, as well as for details underlying the calculations below.  The key point here is that these states, belonging to the Hilbert space of the fundamental theory, offer a unique opportunity to tackle the problem of the emergence of cosmological dynamics directly at the level of a candidate quantum gravity theory, taking into account all degrees of freedom.

\section{Group Field Theory}
\label{s.gft-phi}

Group field theory can be understood as a second-quantized reformulation of LQG \cite{Oriti:2013aqa}, where the field operators create and annihilate quanta of geometry corresponding to spin network nodes.  To be specific, in GFT models for geometry coupled to a (massless) scalar field the elementary GFT field operators, in the spin representation, are
\be
\h\vp^{j_v, \iota}_{m_v}(\phi) = \h\vp^{j_1, j_2, j_3, j_4; \iota}_{m_1, m_2, m_3, m_4}(\phi)
\ee
and its conjugate $\h\vp^\dag{}^{j_v, \iota}_{m_v}(\phi)$ \cite{Oriti:2013aqa, Oriti:2016qtz}.  Here for simplicity (and as suggested by LQC) we are only considering four-valent spin network nodes, this can be seen by the four $j_i$ and $m_i$ arguments in the field operator.  The field operators can be taken to be bosonic ladder operators acting on a Fock space. The creation operator $\h\vp^\dag{}^{j_v, \iota}_{m_v}(\phi)$ creates a quanta of geometry: a four-valent spin network node, each link coloured by $j_i$ and $m_i$, and with the intertwiner $\iota$ associated to the usual gauge invariance of spin network nodes, for a given value $\phi$ of the scalar field.  

The familiar geometric operators of LQG can be imported to the second-quantized language; generic operators can be constructed in terms of strings of ladder operators.  The presence of an additional argument, the matter field $\phi$, also permits the definition of relational observables, i.e., operators evaluated at a specific value of $\phi$, and these are crucial for describing physical evolution in a fully diffeomorphism invariant language.  For example, the number operator counting the number of quanta above the Fock vacuum $|0\rangle$ when the matter field assumes the value $\phi$ is given by
\be
\h N(\phi) = \sum_{j, m, \iota} (\h\vp^\dag)^{j_v, \iota}_{m_v}(\phi) \: \h\vp^{j_v, \iota}_{m_v}(\phi).
\ee

The microscopic dynamics is controlled by equations of motion that also determine the Feynman expansion.  The various monomials in the field operators are chosen to allow for a natural mapping between Feynman graphs and four-dimensional (simplicial) complexes decorated with geometric data, with amplitudes encoding suitably discretized gravitational actions evaluated on the given discrete configurations. In particular, one can define GFT models where these amplitudes exactly match any desired spin foam amplitudes (e.g., those directly motivated from LQG).  These are easily generated starting from simple action functionals, that we split into linear and non-linear parts as $S[\vp, \b\vp] = K[\vp, \b\vp] + V[\vp, \b\vp]$, with the kinetic term encoding the edge amplitude of the spin foam model and having the form (with a minimally coupled massless scalar field)
\begin{align} \label{kinetic}
K = \sum_{j, m, \iota} \int \dd\phi_1 \dd\phi_2
\bigg[ & \b\vp^{j_{v_1}, \iota_1}_{m_{v_1}}(\phi_1) \:
\vp^{j_{v_2}, \iota_2}_{m_{v_2}}(\phi_2) \nn \\ & \: \times
K^{j_{v_1}, j_{v_2}, \iota_1, \iota_2}_{m_{v_1}, m_{v_2}}((\phi_1 - \phi_2)^2) \bigg],
\end{align}
while the potential $V[\vp, \b\vp]$ encodes the vertex amplitude. $V[\vp, \b\vp]$ is (for simplicial GFT models) of fifth order in the classical GFT field variables $\vp$ and $\b\vp$ (i.e., the classical fields corresponding to the field operators $\h\vp$ and $\h{\b\vp}$ in the quantum theory), and is local in the scalar field $\phi$.

It is convenient to rewrite the kinetic term as a derivative expansion in $\phi$ in the field variable $\vp^{j_{v_2}, \iota_2}_{m_{v_2}}(\phi_2)$ around $\phi_2 = \phi_1 = \phi$, giving
\be
K = \sum_{n=0}^\infty \sum_{j, m, \iota} \int \!\! \dd\phi \,
\b\vp^{\, j_{v_1}, \iota_1}_{m_{v_1}}\!(\phi) \,
\partial_\phi^{2n} \vp^{j_{v_2}, \iota_2}_{m_{v_2}}\!(\phi)
(K^{(2n)})^{j, \iota}_m,
\ee
where the notation on $K^{j, \iota}_m$ has been compressed, and
\be
(K^{(2n)})^{j, \iota}_m = \int \dd u \: \f{u^{2n}}{(2n)!} K^{j, \iota}_m(u^2).
\ee
In cases where the difference between $\phi_1$ and $\phi_2$ in \eqref{kinetic} is small compared to the Planck mass, a good approximation to the full kinetic term can be provided by a truncation of the derivative expansion.  This is expected to be the case here since the momentum of a massless scalar field in an FLRW space-time is a constant of the motion typically well below the Planck scale, and furthermore this is a natural approximation in a hydrodynamical setting where the fluid density (i.e., $|\vp^{j,\iota}_m(\phi)|^2$) is expected to vary slowly with respect to its arguments.  Therefore, here we will only keep the first two terms $n=0, 1$.

Finally, for a GFT model with the action $S[\vp, \b\vp]$, the quantum equations of motion for a state $|\Psi\ket$ are simply
\be \label{qeom}
\wh{ \f{\de S}{\de \b\vp} } \, |\Psi\ket = 0,
\ee
together with the conjugate of this equation.

As with any interacting field theory, it is not known how to obtain the general solution of these equations. The particular formulation given by GFT, however, allows us to make use of ideas and methods that are used in analogous problems in condensed matter physics. We will seek some state that approximates a full solution state $| \Psi \rangle$, at least for a restricted set of observables.  The restriction to the case of homogeneous cosmologies suggests that these states should be modeled with a wave function homogeneity principle \cite{Gielen:2013kla, Gielen:2013naa, CondensatesReview, Oriti:2015qva}, i.e., by condensate states in which the wave functions associated to each of the quanta are the same.

\section{Isotropic Condensate States}
\label{s.cond}

The simplest way to model such cosmological states, including an arbitrary large number of quanta, is to use the field coherent states
\be \label{cond}
|\sigma\ket = e^{-\|\sigma\|^2/2} \exp \left( \sum_{j, m, \iota} \sigma^{j_v, \iota}_{m_v}(\phi) 
(\h\vp^\dag)^{j_v, \iota}_{m_v}(\phi) \right) | 0 \ket,
\ee
where $\sigma^{j_v, \iota}_{m_v}(\phi)$ is the condensate wave function and $\|\sigma\|^2 = \int \dd\phi \: \|\sigma(\phi)\|^2$.  An important point here is that the condensate wave function is not normalized: rather the norm of $\sigma^{j_v, \iota}_{m_v}(\phi)$,
\be
\|\sigma(\phi)\|^2 = \sum_{j, m, \iota} |\sigma^{j_v, \iota}_{m_v}(\phi)|^2,
\ee
is the expectation value of the number operator $\h N(\phi)$ on the condensate state $|\sigma\ket$ at the relational time $\phi$.

These states have been extensively studied in the GFT context \cite{Gielen:2013kla, Gielen:2013naa, CondensatesReview} as approximate solutions of the quantum equations of motion. As they neglect correlations between different quanta (and thus the connectivity of the spin network nodes), these are approximate solutions only in regimes in which the interaction term in \eqref{qeom} is subdominant.  Note that it is reasonable to discard the connectivity information for the spatially flat FLRW space-time since (i) the connectivity information is not needed in order to extract the key geometrical observable, i.e., the total volume, and (ii) there is no need to encode any spatial curvature in the GFT condensate.  For more complex space-times, it may no longer be possible to disregard the connectivity information.

Since we are only interested in the homogeneous and isotropic degrees of freedom, it is possible to choose a particularly simple form of the condensate wave function by imposing that the condensate wave function be isotropic, i.e., that all of the spin labels be equal, and that the other geometric indices be uniquely defined by $j$.  Hence, for an isotropic condensate wave function,
\be
\sigma^{j_v, \iota}_{m_v}(\phi) = C^{j_v, \iota}_{m_v} \cdot \sigma_j(\phi),
\ee
where the $C^{j_v, \iota}_{m_v}$ are uniquely determined by the value of $j$ (in particular, the intertwiner is chosen so that it is an eigenvalue of the LQG volume operator and that its eigenvalue is the largest possible for a spin network node with four links all coloured by $j$, see \cite{Oriti:2016qtz} for details).  Therefore, the coarse-grained degrees of freedom of isotropic GFT condensate states are entirely captured by the functions $\sigma_j(\phi)$, one for each spin.

We stress that these additional conditions and approximations on the trial states are necessary to extract cosmological physics at a coarse-grained level.  Generic GFT or LQG states will not correspond to any continuum space-times; and while the isotropic condensate states are a reasonable ansatz for FLRW space-times, this can only be determined by analyzing the resulting effective dynamics for coarse-grained observables like the total spatial volume.

The dynamics for the condensate states \eqref{cond} are obtained by asking that they approximately solve the quantum equations of motion \eqref{qeom}.  To be specific, we assume a Gross--Pitaevskii form of the dynamics corresponding to the expectation value of the equations of motion,
\be \label{cond-eom}
\bra \sigma | \, \wh{ \f{\de S}{\de \b\vp} } \, | \sigma \ket = 0,
\ee
which is of course a weaker condition than \eqref{qeom}.

For the isotropic GFT condensate states \eqref{cond}, and for a GFT model with a minimally coupled massless scalar field whose geometric contribution is based on the Engle--Livine--Pereira--Rovelli spin foam model \cite{Engle:2007wy} (the most developed one for 4D Lorentzian quantum gravity), \eqref{cond-eom} gives the equation of motion for the $\sigma_j(\phi)$
\be \label{eom-sigma}
A_j \partial_\phi^2 \sigma_j(\phi) - B_j \sigma_j(\phi) + w_j \b\sigma_j(\phi)^4 = 0.
\ee
It is clear that the scalar field $\phi$ is acting as a relational clock here and can be interpreted as `time'.  This will be important when extracting the coarse-grained cosmological dynamics from this condensate state.  Here $A_j$ depends on a combination of $(K^{(2)})^{j, \iota}_m$ and $C^{j, \iota}_m$, $B_j$ depends on $(K^{(0)})^{j, \iota}_m$ and $C^{j, \iota}_m$, while $w_j$ (and the form of the fourth order term in the condensate wave function) depends on the specific form of $V[\vp, \b\vp]$ as well as $C^{j, \iota}_m$.  We refer to \cite{Oriti:2016qtz} for the exact relations.

This Gross--Pitaevskii approximation of the dynamics, based on the ansatz \eqref{cond} for the quantum state, is expected to hold in a weakly interacting regime where the non-linear term in \eqref{eom-sigma} is subdominant.  (Note that the simplicity constraints encoded in the potential in the GFT action \emph{are} present and so the geometric interpretation of the GFT quanta is justified.  Here we consider the regime where the contribution from the potential to the equations of motion is subdominant, not vanishing.  As an aside, this condition is clearly necessary for explicit calculations in the spin foam perturbative expansion to be meaningful.)  Indeed, when the last term in \eqref{eom-sigma} will become important for sufficiently large $|\sigma_j(\phi)|$, this approximation will no longer be valid \cite{Oriti:2016qtz}. In the `strong interaction' regime, correlations between quanta (encoding connectivity information) will play a prominent role, and simple coherent states will have to be replaced by more elaborated condensate states (perhaps of the type introduced in \cite{Oriti:2015qva}).  Therefore, in the following we will only consider the mesoscopic regime in which there are a greater number of quanta than some reasonable minimum $N_{min}$ for which the hydrodynamical picture is valid (i.e., $\sum_j|\sigma_j|^2 \gg N_{min}$), but small enough for the non-linear term in \eqref{eom-sigma} to be subdominant compared to the linear terms (i.e., $\forall \, j,~|\sigma_j| \ll |B_j / w_j|^{1/3})$. Considerations on the estimated size of fluctuations in the volume observable, experience with real condensates, and arguments relating GFT condensates to continuum geometries, together suggest $N_{min} \gtrsim 10^{3}-10^4$ \cite{CondensatesReview}. Detailed calculations with explicit solutions of the effective cosmological GFT dynamics are needed to obtain a more precise estimate.  The existence of such a regime depends on the specific GFT model via its action (and in particular the relative amplitudes of the prefactors to the kinetic and interaction terms) and can be determined by checking the solutions to the full non-linear equations of motion.  For a sufficiently small interaction term in the GFT action such a regime is guaranteed to exist, and the smaller the interaction term is the longer this mesoscopic regime will exist; some specific (phenomenological) examples have recently been studied in \cite{deCesare:2016rsf} that clearly show the existence of this mesoscopic regime for some GFT actions.  We will restrict our attention in the following to the GFT models where such a mesoscopic regime exists.

In this mesoscopic regime, rewriting $\sigma_j(\phi) = \rho_j e^{i \theta_j}$ in terms of its modulus and phase, and denoting derivatives with respect to $\phi$ by primes, \eqref{eom-sigma} gives the two equations of motion
\be \label{eom-rho}
\rho_j'' - \Big( m_j^2 + (\theta_j')^2 \Big) \rho_j \approx 0,
\ee
\be \label{eom-theta}
\rho_j \theta_j'' + 2 \rho_j' \theta_j' \approx 0,
\ee
with $m_j^2 = B_j / A_j$.
In this approximation, the quantities
\be
E_j = (\rho_j')^2 + \rho_j^2 (\theta_j')^2 - m_j^2 \rho_j^2,
\ee
\be
Q_j = \rho_j^2 \theta_j',
\ee
are conserved with respect to the relational time $\phi$.  Then, using $Q_j$, \eqref{eom-rho} becomes
\be \label{eom-rho-q}
\rho_j'' - m_j^2 \rho_j - \f{Q_j^2}{\rho_j^3} \approx 0.
\ee

\section{Emergent Friedmann Equations}
\label{s.fried}

To extract the dynamics of the large-scale coarse-grained cosmological observables, these observables must be related to the microscopic GFT degrees of freedom.  In this case, this procedure is straightforward since the quantities of interest are the total volume (corresponding to the cube of the scale factor) and the momentum of the massless scalar field, evaluated at a (relational) time $\phi$.

These operators, evaluated on the isotropic GFT condensate states, are respectively
\be \label{vol}
V(\phi) = \sum_j V_j \, \rho_j(\phi)^2,
\ee
where, given the definition of the isotropic condensate states, $V_j \sim j^{3/2} \lp^3$ is the largest eigenvalue of the LQG volume operator for a spin network node with four links labeled by $j$, and
\be
\pi_\phi(\phi) = \f{\hbar}{2 i} \sum_j \Big( \b\sigma_j(\phi) \sigma_j(\phi)'
- \b\sigma_j(\phi)' \sigma_j(\phi) \Big).
\ee
It is easy to check that $\pi_\phi = \hbar \sum_j Q_j$ and that therefore the momentum of the massless scalar field is a constant of the motion in this context.  This is exactly the continuity equation: for an FLRW space-time with a massless scalar field, the continuity equation reduces to $\pi_\phi' = 0$.

Therefore, the only other equation of motion to be recovered in order to determine the coarse-grained cosmological dynamics is the Friedmann equation relating the Hubble rate to the energy density of the matter field.  This can be calculated, using $V' = 2 \sum_j V_j \, \rho_j \rho_j'$, to be
\be \label{gen-friedmann}
\left( \f{V'}{3V} \right)^2 = \left( \f{2 \sum_j \! V_j \rho_j {\rm sgn}(\rho_j')
\sqrt{E_j - Q_j^2 / \rho_j^2 + m_j^2 \rho_j^2 \,} }{3 \sum_j V_j \rho_j^2} \right)^2.
\ee
This and $\pi_\phi'=0$ are the emergent (generalized) Friedmann dynamics coming from the microscopic GFT quantum equations of motion for isotropic condensate states.

This can be compared to the classical Friedmann equation of general relativity with a massless scalar field, in which case the energy density is $\rho = \pi_\phi^2 / 2 V^2$ (with $\pi_\phi$ constant), and using the scalar field as a relational clock is the same as choosing the lapse to be $N = V / \pi_\phi$, giving
\be \label{fr-class}
\left( \f{V'}{3V} \right)^2 = \f{4 \pi G}{3}.
\ee
The usual Hubble rate is given by $H = \pi_\phi V' / 3 V^2$.

In order to check that \eqref{gen-friedmann} gives the correct classical limit, we take $\rho_j$ to be sufficiently large so that $m_j^2 \rho_j^2$ is the dominant term in the square root in the numerator.  This is a large volume limit, which in FLRW space-times also corresponds to the low-curvature classical limit.  Then, in this limit
\be
\left( \f{V'}{3V} \right)^2 = \left( \f{2 \sum_j \! V_j m_j \rho_j^2 }{3 \sum_j V_j \rho_j^2} \right)^2.
\ee
It is clear that a sufficient (although not necessary) condition to reproduce the classical Friedmann equation \eqref{fr-class} is%
\footnote{The problem of including a scalar field as matter content in a GFT model (beyond the general framework presented here) remains open.  Recovering the correct Friedmann equation in the classical limit provides a simple constraint for the action of such a GFT model, which initially has the same ambiguities as spin foam models or indeed any discrete gravity model.}
$m_j^2 = 3 \pi G$ for all $j$.  This shows that for GFT models where the parameters in the action are chosen such that $m_j^2 = 3 \pi G$, the coarse-grained, large-scale dynamics of isotropic condensate states are modified Friedmann equations with the correct classical limit, and with quantum gravity corrections when the space-time curvature is sufficiently large%
\footnote{For the spatially flat FLRW space-time, the space-time curvature is entirely determined by the trace of the extrinsic curvature, which can be calculated from the dynamics of the spatial volume in terms of the relational time $\phi$.  Therefore, in this special case, the space-time curvature is easy to calculate from $|\sigma\ket$.}.
This is our first main result, and is in agreement (although in a more general form as well as more closely grounded in a fundamental GFT model) with previous work on GFT condensates \cite{Gielen:2013kla, Gielen:2013naa}.

This result also suggests that the classical Friedmann dynamics is a rather universal emergent feature of GFT quantum gravity in the regime considered here, since most differences between specific microscopic models can be encoded in the interaction terms which are here subdominant. This is consistent with the hydrodynamics perspective on cosmology we have adopted.

Importantly, the formalism developed here also allows one to go beyond earlier results studying GFT condensates and to study the effect of quantum gravity corrections on the dynamics.  The quantum gravity corrections play an important role when the space-time curvature becomes large and give rise to a striking feature of the emergent cosmological dynamics. As is obvious from \eqref{eom-rho-q}, the $\rho_j$ can never become zero but will instead necessarily reach a turning point due to the repulsive and divergent `potential' $-Q_j^2 / \rho_j^3$, assuming $Q_j \neq 0$.  Note that in an FLRW space-time, $\pi_\phi \neq 0$ and so at least one $Q_j$ must be non-zero.  Therefore, it immediately follows that at least one $\rho_j > 0$ for all $\phi$, and clearly the volume $V$ will never be zero.  As a result, the big-bang and big-crunch singularities of classical general relativity are generically resolved here, due to quantum corrections, within a full quantum gravity formalism.  Furthermore, there will be a bounce since there is only one turning point for each $\rho_j$ in \eqref{eom-rho-q}. This is our second main result.

Finally, it is possible to consider the case where $\sigma_j(\phi)$ only has support on a single $j=j_o$---these are GFT states that closely match the heuristic relation between LQG and LQC \cite{Ashtekar:2006wn, Bojowald:2007ra, Ashtekar:2009vc, Pawlowski:2014nfa}.  In that case, the sums in the Friedmann equation \eqref{gen-friedmann} trivialize, and (setting $m_{j_o}^2 = 3 \pi G$)
\be
\left( \f{V'}{3V} \right)^2 = \f{4 \pi G}{3} \left( 1 - \f{\rho}{\rho_c} \right) + \f{4 V_{j_o} E_{j_o}}{9 V},
\ee
where $\rho = \pi_\phi^2 / 2 V^2$ is the energy density of the massless scalar field and the critical energy density is $\rho_c = 3 \pi G \hbar^2 / 2 V_{j_o}^2$, with $V_{j_o} \sim j_o^{3/2} \lp^3$. Interestingly, this equation is almost identical to the LQC effective equations, with the only difference being the last term.  For states with $E_{j_o} = 0$, the resulting effective Friedmann dynamics will then be exactly that of LQC.  While the geometric interpretation of the `energy' $E_{j_o}$ is not clear, its effect on the dynamics is: if $E_{j_o} > 0$, the cosmological bounce will occur at a higher space-time curvature, while if $E_{j_o} < 0$ then the bounce will occur at a lower space-time curvature. Even the effective LQC dynamics, then, can be obtained from a complete quantum gravity formalism. This is our third main result.

\section{Discussion}
\label{s.disc}

In this paper, we have shown how cosmological dynamics emerge from a family of states in full quantum gravity.  This goes beyond the minisuperspace models of quantum cosmology theories, including LQC, which are based on quantizing only the degrees of freedom of a symmetry-reduced space-time, rather than starting from a quantum theory of gravity with all of its degrees of freedom. These results open a very promising route for studying cosmology from within a complete quantum gravity formalism, and show that the macroscopic cosmological consequences of the underlying microscopic quantum geometry can be analysed in detail.

More specifically, we have argued that cosmological states in GFT (for models with a direct LQG interpretation of quantum states and dynamics) correspond to highly-excited (with respect to the degenerate vacuum) condensate states, and showed how to extract the hydrodynamical equations for the coarse-grained cosmological observables from the microscopic quantum dynamics.  The correct Friedmann equations are recovered in the classical limit for a large class of GFT models (i.e., those whose action gives $m_j^2 = 3 \pi G$).

Strikingly, the cosmological singularity is generically resolved, due to quantum gravity corrections, and is replaced by a bounce. Finally, single-spin condensate states reproduce very closely the effective LQC dynamics.  Of course, it is also possible to consider states that have support on many spins, and the dynamics of these states are given by a generalization of the LQC effective dynamics.

An important challenge now is to develop a framework for cosmological perturbation theory in GFT.  This will make it possible to calculate quantum gravity effects on scalar and tensor perturbations, and compare these predictions---coming from full LQG/GFT---to observations of the cosmic microwave background.

The results in this paper are based on a number of necessary assumptions: (i) we assumed that, in the cosmological sector of GFT and LQG, the massless scalar field does not evolve rapidly compared to the Planck mass, keeping only the first two terms of the derivative expansion of the GFT kinetic term, (ii) we neglected the connectivity of the spin networks since it is not needed in order to extract isotropic and homogeneous observables, (iii) we only imposed that the expectation value of the quantum equations be zero, and (iv) we worked in the regime in which the contribution of interactions to the equations of motion are subdominant, this being necessary for the approximation (ii) to be valid.

Despite the above limitations, these results provide important insights on the cosmological sector of GFT and LQG, and are also in strong qualitative agreement with those obtained in the (homogeneous) LQC context.

\acknowledgments

We thank the participants of the ``Cosmology from Quantum Gravity'' workshop at the AEI, in particular M. Sakellariadou, M. Bojowald, J. Mielczarek, S. Gielen and A. Pithis, for helpful discussions.
This work was partially supported by the John Templeton Foundation with grant PS-GRAV/1401.

\raggedright

\end{document}